\documentclass[11pt]{article}

\usepackage{eepic,epic}

\makeatletter%
\def\nottoobig#1{{\hbox{$\left#1\vcenter to1.111\ht\strutbox{}\right.\n@space$}}}
\makeatother%

\makeatletter%
\def\@begintheorem#1#2{\trivlist\item[\hskip\labelsep{\bf #1\ #2}]}
\makeatother
\makeatletter %

\newlength{\filength}
\newsavebox{\gcbox}
\sbox{\gcbox}{\framebox[\filength]{\rule{0ex}{2ex}}}

\newlength{\leftjustindent}
\newlength{\@leftjustindent}
\def\leftjust{\let\\\@leftjustcr\let\end\@endleftjust
  \addtolength{\@leftjustindent}{\leftjustindent}
  \vcenter\bgroup
  \halign\bgroup
    \hbox to\displaywidth{
      \rule{\@leftjustindent}{0ex}$\displaystyle##$\hfill
      }\crcr
}
\def\endleftjust{\crcr\egroup\egroup\endgroup}
\def\@endleftjust#1{\crcr\egroup\egroup\@checkend{#1}\endgroup}
\def\@leftjustcr{\crcr}

\newtheorem{theorem}{Theorem}[section]

\newcommand{\qedblob}{\mbox{\rule[-1.5pt]{5pt}{10.5pt}}}
\def\literalqed{{\ \nolinebreak\hfill\mbox{\qedblob\quad}}}

\def\qed{\literalqed}

\newtheorem{lemma}[theorem]{Lemma}

\newcommand{\singlespacing}{\let\CS=
\@currsize\renewcommand{\baselinestretch}{1}\tiny\CS}
\newcommand{\singlespacingplus}{\let\CS=
\@currsize\renewcommand{\baselinestretch}{1.25}\tiny\CS}
\newcommand{\doublespacing}{\let\CS=
\@currsize\renewcommand{\baselinestretch}{1.75}\tiny\CS}
\newcommand{\draftspacing}{\let\CS=
\@currsize\renewcommand{\baselinestretch}{2.0}\tiny\CS}

\makeatother%

\newtheorem{definition}[theorem]{Definition}

\makeatletter
\clubpenalty=\@highpenalty
\widowpenalty=\@highpenalty
\makeatother

\makeatletter
\newcommand{\niceonespacing}{\let\CS=\@currsize\renewcommand{\baselinestretch}{1.1}\tiny\CS}\newcommand{\nicetwospacing}{\let\CS=\@currsize\renewcommand{\baselinestretch}{1.2}\tiny\CS}
\newcommand{\nicethreespacing}{\let\CS=\@currsize\renewcommand{\baselinestretch}{1.3}\tiny\CS}
\newcommand{\singlespacingplusplus}{\let\CS=\@currsize\renewcommand{\baselinestretch}{1.35}\tiny\CS}
\newcommand{\nicefivespacing}{\let\CS=\@currsize\renewcommand{\baselinestretch}{1.5}\tiny\CS}
\newcommand{\nicesixspacing}{\let\CS=\@currsize\renewcommand{\baselinestretch}{1.6}\tiny\CS}
\newcommand{\nicefourspacing}{\let\CS=\@currsize\renewcommand{\baselinestretch}{1.4}\tiny\CS}
\newcommand{\nicefoospacing}{\let\CS=\@currsize\renewcommand{\baselinestretch}{1.05}\tiny\CS}
\makeatother

\makeatletter
\def\@cite#1#2{[#1\if@tempswa , #2\fi]}
\makeatother

\makeatletter
\def\@citex[#1]#2{\if@filesw\immediate\write\@auxout{\string\citation{#2}}\fi
  \def\@citea{}\@cite{\@for\@citeb:=#2\do
    {\@citea\def\@citea{,\linebreak[0]}\@ifundefined
       {b@\@citeb}{{\bf ?}\@warning
       {Citation `\@citeb' on page \thepage \space undefined}}%
\hbox{\csname b@\@citeb\endcsname}}}{#1}}
\makeatother

\newcommand{\sharpp}{{\rm \#P}}

\newcommand{\p}{{\rm P}}
\newcommand{\littlep}{{\rm p}}

\newcommand{\npnp}{{\rm NP^{NP}}}
\newcommand{\bh}{{\rm BH}}
\newcommand{\BH}{{\rm BH}}

\newcommand{\np}{{\rm NP}}

\newcommand{\pp}{{\rm PP}}

\newcommand{\conp}{{\rm coNP}}

\newcommand{\pij}{{\p^{\bh_i:\bh_j}}}
\newcommand{\pji}{{\p^{\bh_j:\bh_i}}}

\newcommand{\sigmatwo}{{\Sigma_2^{\littlep}}}
\newcommand{\sigmak}{{\Sigma_k^{\littlep}}}
\newcommand{\sigmai}{{\Sigma_i^{\littlep}}}
\newcommand{\sigmaj}{{\Sigma_j^{\littlep}}}

\newcommand{\ph}{{\rm PH}}

\newcommand{\superseteq}{\supseteq}

\newcommand{\substar}{\mbox{\tiny$<$}}
\newcommand{\superstar}{\mbox{\tiny$>$}}

\nicefoospacing
\newcommand{\manyone}{\mbox{$\,\leq_{\rm m}^{{\littlep}}$\,}}
\newcommand{\Turing}{\mbox{$\,\leq_{\rm T}^{{\littlep}}$\,}}

\newcommand{\calc}{\ensuremath{{\cal C}}}
\newcommand{\cald}{\ensuremath{{\cal D}}}

\newcommand{\condition}{\,\nottoobig{|}\:}
\def\land{{\,\, \wedge \;}}

\begin{document}

\bibliographystyle{alpha}

\title{An Introduction to Query Order\protect\thanks{%
\protect\singlespacing
Supported in part 
by grants
NSF-CCR-9322513
and
NSF-INT-9513368/\protect\linebreak[0]DAAD-315-PRO-fo-ab.}}

\author{
Edith Hemaspaandra\footnote{
\protect\singlespacing
{\tt edith@bamboo.lemoyne.edu}.
Department of Mathematics, 
Le Moyne College,
Syracuse, NY 13214, 
USA\@.
Work done in part while 
visiting Friedrich-Schiller-Universit\"at Jena.} 
\and
Lane A. Hemaspaandra\footnote{
\protect\singlespacing
{\tt lane@cs.rochester.edu}.
Department of Computer Science,
University of Rochester,
Rochester, NY 14627, 
USA\@.
Work done in part while 
visiting Friedrich-Schiller-Universit\"at Jena.} 
\and
Harald Hempel\footnote{
\protect\singlespacing
{\tt hempel@informatik.uni-jena.de}.
Institut f\"ur Informatik,
Friedrich-Schil\-ler-Univer\-si\-t\"at Jena,
07740 Jena, 
Germany.  Work done in part while 
visiting Le Moyne College.}
}

\makeatletter%

\newcount\hour  \newcount\minutes  \hour=\time  \divide\hour by 60
\minutes=\hour  \multiply\minutes by -60  \advance\minutes by \time
\def\mmmddyyyy{\ifcase\month\or Jan\or Feb\or Mar\or Apr\or May\or Jun\or Jul\or
  Aug\or Sep\or Oct\or Nov\or Dec\fi \space\number\day, \number\year}
\def\hhmm{\ifnum\hour<10 0\fi\number\hour :%
  \ifnum\minutes<10 0\fi\number\minutes}
\def\Draft{{\it Draft of \mmmddyyyy}}
\makeatother%

\date{August 29, 1997}

{\singlespacing

\singlespacing\maketitle

}

\nicefoospacing

{

\noindent{\bf Abstract:} \quad
Hemaspaandra, Hempel,
and Wechsung
\cite{hem-hem-wec:tSPECIALwithJ:query-order-bh} 
raised the following questions:
If one is allowed one question to each of two different information sources, 
does the order in which one asks the questions affect the class of problems 
that one can solve with the given access? If so, which order yields the 
greater computational power?

The answers to these questions have been 
learned---inasfar as they can
be learned without resolving whether or not the polynomial hierarchy
collapses---for both the polynomial hierarchy and the boolean
hierarchy.  In the polynomial hierarchy, query order never matters.
In the boolean hierarchy, query order sometimes does not matter and,
unless the polynomial hierarchy collapses, sometimes does matter.
Furthermore, the study of query order has yielded dividends in
seemingly unrelated areas, such as bottleneck computations and
downward translation of equality.  

In this article, we present some of the central results on query
order.  The article is written in such a way as to encourage the
reader to try his or her own hand at proving some of these results.
We also give literature pointers to the quickly growing set of
related results and applications.

} %

\section{Introduction}

So, you're at a theory conference and the coming session 
strikes you as potentially boring. 
You walk into the lobby in search of more coffee and some theoretical 
chit-chat, and you get more than you bargained for. 
Poof! 
A well-dressed stranger appears seemingly from nowhere. 
His name tag is hidden under a lapel. 
Under his arm is a stack of books.
A second edition of Garey and Johnson?
{\em The Polynomial Hierarchy Does Not Collapse\/} by M.~Sipser and 
A.~Yao?
Who is this guy?
You don't have a clue.
Wait a moment...~is that Volume~4 of Knuth under his arm?
Aha!
Clearly, this is a time-traveler from the distant future!
\medskip

\noindent{\bf Stranger:}\quad 
These books will all be yours if you can solve 
the following small puzzle.
Here are two black boxes.
One is marked SAT, and in unit time it solves satisfiability 
(using magical means not widely available in the 20th century).
The other black box
is marked IEI, and in unit time it solves the well-known 
$\npnp$-complete problem IntegerExpressionInequivalence 
(again using magic).
I'll ask you to write a polynomial-time algorithm that accepts the 
language FOO, and that on each input makes at most one query to each of 
the black boxes. 
Actually, at this moment I don't demand that you write the algorithm.
I just want you to decide on a fixed order in which you will access the 
black boxes.
Either you must (right now) commit to accessing SAT first on each input, 
or you must (right now) commit to accessing IEI first on each input.

\noindent {\bf You:}\quad  Well, before I commit, will you tell me 
a bit about what problem FOO is?

\noindent{\bf Stranger:}\quad No.

\noindent {\bf You:}\quad
Oh. Wait a second. You're trying to trick me. 
$\npnp$ 
(as captured by its complete
set,~IEI)
is {\em much\/} more powerful than $\np$
(as captured by its complete set,~SAT). So for both classes 
you are speaking of the query to~SAT is superfluous. 
That is: $\p^{{\rm IEI} [1]}=\p^{\rm IEI:SAT}=\p^{\rm SAT:IEI}$, 
where ``$[1]$'' 
indicates one query and $\p^{ A : B}$ indicates a $\p$ machine 
making at most 
one query to (the set)~$A$ set followed by at most one query to (the 
set)~$B$.

\noindent{\bf Stranger:}\quad 
False! Your reasoning is tempting, but is 
not valid.
In fact, that weak SAT query provably gives strictly more computational 
power, as you would know if you had read a certain 
{\em BEATCS\/} 
article~\cite{hem-hem-hem:jSPECIALJOKEasifappearedAlready:query-order-survey}
and had applied it in light of the Sipser-Yao book under my arm.

\noindent {\bf You:}\quad
I've got you now, as I {\em did\/} read that obscure 
little article!  
I retract the $\p^{\rm IEI [1]}=\p^{\rm IEI:SAT}$ claim I just 
made, but I reassert the $\p^{\rm IEI:SAT}=\p^{\rm SAT:IEI}$ claim! 
Order does not matter here, and so either order I commit to now will leave 
it equally likely that~FOO can be solved via the given access order!

\noindent{\bf Stranger:}\quad Sheesh...~I thought no one except the
authors and the editor had read that column. 
Well, you've beaten me so here are your books.
 
\noindent {\bf You:}\quad
But this Sipser-Yao book is from the year 2010. 
If I just steal their 
result and proof, and publish them now, won't that change the future 
and cause the book never to be written, in which case how can I be meeting 
you and receiving this book in the first place?

\noindent{\bf Stranger:}\quad
(Suddenly starts fading away, but before he disappears 
you hear)~~
I commend to you the 
{\small Nebula-Award-winning book
{\em Timescape}}~{\footnotesize{}\cite{ben:b:timescape}, 
which also has quite a bit to say about}
{\scriptsize curiosity-driven research, }
{\tiny scientific ethics, and...}

\bigskip\indent
You suddenly feel a bit disoriented and---distractedly tossing the books into 
a trash bin and mumbling about swearing off that ninth cup of coffee---you go 
back into the lecture hall and listen with
half-hearted attention to an
in-progress lecture on query order.

\section{Query Order in the Polynomial Hierarchy}
\label{s:PH}

\subsection{Results}
\label{subs:PH-r-o}

Recall from the introduction that,
for any sets $A$ and $B$,
$\p^{A:B}$ denotes the class 
of languages that can be accepted via~P machines making at most one 
query to~$A$ {\em followed by\/} at most one query to~$B$.
Similarly, for any classes $\calc$ and $\cald$, 
$\p^{\calc:\cald}$ denotes the class 
of languages that can be accepted via~P machines making at most one 
query to a set from~$\calc$ {\em followed by\/} at most one query to
a set from~$\cald$.  That is, 
$$\p^{\calc:\cald} = \bigcup_{C\in \calc,\,D\in\cald\,}\nolinebreak
\p^{C:D}.$$
These notions and notations were introduced by Hemaspaandra, Hempel, and 
Wechsung~\cite{hem-hem-wec:tSPECIALwithJ:query-order-bh}, who studied them 
for the 
case in which $\calc$ and $\cald$ are levels of the boolean hierarchy 
(see Section~\ref{s:BH}); in brief, they proved that query order usually 
does matter in the boolean hierarchy.

Hemaspaandra, Hemaspaandra, and 
Hempel~\cite{hem-hem-hem:tREVISEDmoreRecentThanFCTasHasLeafStuff:query-order-ph,hem-hem-hem:ctoappear:query-order-ph},
following up on the concept of query order, asked whether query order also 
matters in the polynomial hierarchy.
What they found was that query order {\em never\/} matters in the polynomial 
hierarchy. 
This is stated formally below as
Theorem~\ref{t:exchange}. 

\begin{definition}~\cite{sto:j:poly}
\begin{enumerate}
\item $\Sigma_0^{\littlep}=\p$. 
\item For each 
$k>0$, $\Sigma_k^{\littlep}=\np^{\Sigma_{k-1}^{\littlep}}$.
(For example, 
$\Sigma_1^{\littlep}=\np$ and 
$\Sigma_2^{\littlep}=\np^{\rm NP}$.)

\item $\ph=\bigcup\limits_{k\geq 0}\Sigma_k^{\littlep}$.
\end{enumerate}
\end{definition}

\begin{theorem}\label{t:exchange}~\cite{hem-hem-hem:ctoappear:query-order-ph}
\quad
For each $i,j \geq 0$, 
$$\p^{\sigmai:\sigmaj}=\p^{\sigmaj:\sigmai}.$$
\end{theorem}

In fact, in all but the ``diagonal'' cases of this theorem (where order elimination is impossible unless the polynomial hierarchy collapses), one can eliminate order entirely:

\begin{theorem}~\cite{hem-hem-hem:ctoappear:query-order-ph}
\quad
For each $i,j\geq 0$ with $i\not=j$,
$$\p_{1,1\mbox{\scriptsize{-tt}}}^{(\sigmai,\sigmaj)}=\p^{\sigmai:\sigmaj}
=\p^{\sigmaj:\sigmai},$$
where $\p_{1,1\mbox{\scriptsize{-tt}}}^{(\sigmai,\sigmaj)}$ denotes the 
class of 
languages accepted by machines that, in parallel (i.e., simultaneously),
ask at most one question to a $\sigmai$ set and at most one question to a 
$\sigmaj$ set.
\end{theorem}

However, we should now address the ``tempting'' worry you, the reader, 
raised during the introduction.
Let $i<j$. Clearly, $\sigmaj$ can beat $\sigmai$ into a pulp.
For example, it is well known that $\p^{\sigmai} \subseteq 
\sigmaj$.  That is, $\sigmaj$ 
is more powerful than even a polynomially long series of adaptive 
queries to $\sigmai$. 
So it would indeed be tempting to assert:
$\p^{\sigmai:\sigmaj}=\p^{\sigmaj [1]}$ for $i<j$, and indeed there is an 
obvious (but flawed) ``proof'' of this, 
involving having 
$\sigmaj$ simulate the $\sigmai$ query of $\p^{\sigmai:\sigmaj}$, get the 
answer, and then simulate the $\sigmaj$ query of $\p^{\sigmai:\sigmaj}$. 
The flaw is that though $\sigmaj$ 
can do this, it cannot pass to the base P 
machine the information on which truth-table to 
use to process its 
answer; there is a 1-bit information bottleneck!
Indeed, the
tempting 
equality---$\p^{\sigmai:\sigmaj} =
\p^{\sigmaj [1]}$ for $i<j$---is 
outright false unless the polynomial hierarchy 
collapses.
This follows immediately from the more general fact that all ``ordered access 
to the polynomial hierarchy'' classes are either trivially equal or are truly 
different (unless the polynomial hierarchy collapses).

\begin{theorem}~\cite{hem-hem-hem:ctoappear:query-order-ph}
\label{t:mpxx} \quad
Let
$i,j,\ell,m \geq 0$. 
If $\p^{\sigmai:\sigmaj}=\p^{\Sigma_{\ell}^{\littlep}:\Sigma_m^{\littlep}}$, then either 
$\{i,j\}=\{\ell,m\}$ or the polynomial hierarchy collapses.
\end{theorem}

The just-stated theorem merely concludes that the polynomial hierarchy 
collapses (unless $\{i,j\}=\{\ell,m\}$).
In fact, in almost all cases, the polynomial hierarchy collapses to an 
alarmingly low level---one that {\em a priori\/} seems lower than either 
of the classes mentioned in the theorem (this can be seen easily 
from~\cite{hem-hem-hem:jtoappear:downward-translation,buh-for:t:two-queries,hem-hem-hem:t:translating-downwards,hem-hem-hem:ctoappear:query-order-ph}, 
see especially~\cite[Section 3.2]{hem-hem-hem:ctoappear:query-order-ph}).
For example,  
$$\p^{\Sigma_{1997}^{\littlep}:\Sigma_{1999}^{\littlep}}=\p^{\Sigma_{1998}^{\littlep}:\Sigma_{1999}^{\littlep}} 
\Longrightarrow \ph=\Sigma_{1999}^{\littlep},$$
even though {\em a priori\/} one would suspect
that $\Sigma_{1999}^{\littlep}$ is strictly 
contained in 
$\p^{\Sigma_{1997}^{\littlep}:\Sigma_{1999}^{\littlep}}$.

Note that in all the results we have discussed so far, we have a P machine 
doing the querying, i.e., $\p^{\rm SAT:IEI}=\p^{\rm IEI:SAT}$.
In fact, Hemaspaandra et 
al.~\cite{hem-hem-hem:tREVISEDmoreRecentThanFCTasHasLeafStuff:query-order-ph} 
have 
shown that all standard complexity classes 
(in particular, all leaf-definable classes) automatically 
inherit all query-order containments that hold for~P machines.
Thus, for example, since $\p^{\np:\npnp}=\p^{\npnp:\np}$, we may conclude 
immediately that $\pp^{\np:\npnp}=
\pp^{\npnp:\np}$.

\subsection{Proof by Example}\label{s:PH-p}

Our goal here is just to give the general flavor of a proof related to query 
order in the polynomial hierarchy. 
We will prove part of an instance of 
Theorem~\ref{t:exchange}. 
That is, we will partially prove:
$$\p^{\sigmatwo:\np}\subseteq\p^{\np:\sigmatwo}.$$
In particular, we will prove that 
$X\subseteq\p^{\np:\sigmatwo}$, where $X$ is the class of 
languages that are in 
$\p^{\sigmatwo:\np}$ via a 
$\p^{\sigmatwo:\np}$ machine in which the~$\p$ machine 
on each input asks {\em exactly\/} one question to each of 
its oracles, and in which the~$\p$ machine accepts 
if and only if {\em exactly one of its two queries gets
the answer ``yes.''}~~That is, we will do the parity case.

The proof is not hard, and finding it for oneself will help one gain a
feeling for what it is like to study query order.  Thus, we urge the
reader to try to prove this him- or herself before reading the proof
we include below.

\noindent{\bf Proof:}
Let $L \in \p^{\sigmatwo:\np}$.  Let $M$ be a $\p$ machine, 
$A\in \sigmatwo$,  and $B\in \np$ be such that $M$ accepts $L$ and,
on each input,
$M$ makes
exactly one query to $A$ followed by exactly one query to $B$.
(This does not rule out the possibility that were $M(x)$ to 
be given an incorrect answer to its first query it would not 
ask a second query.  However, without loss of generality we can
assume that it always asks exactly one query to each oracle, regardless
of the answer to the first query.  We do assume this both here and in 
the proof of Section~\ref{s:BH-p}.)
We will partially describe a $\p$ machine $M'$ that accepts $L$ with one
query to an $\np$ set followed by one query to a $\sigmatwo$ set.

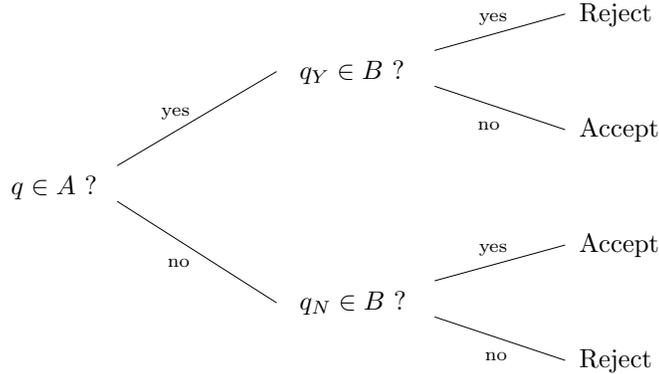
\begin{figure}[tp]
  \begin{center}

\setlength{\unitlength}{0.00050000in}
\begingroup\makeatletter\ifx\SetFigFont\undefined%
\gdef\SetFigFont#1#2#3#4#5{%
  \reset@font\fontsize{#1}{#2pt}%
  \fontfamily{#3}\fontseries{#4}\fontshape{#5}%
  \selectfont}%
\fi\endgroup%
{\renewcommand{\dashlinestretch}{30}
\begin{picture}(6507,3828)(0,-10)
\path(1200,1758)(2850,708)
\path(4500,3333)(5850,3708)
\path(4500,2958)(5850,2508)
\path(4500,933)(5850,1308)
\path(4500,558)(5850,108)
\put(0,1833){\makebox(0,0)[lb]{\smash{{{\SetFigFont{10}{12.0}{\rmdefault}{\mddefault}{\updefault}$~q \in A$~?}}}}}
\put(3000,3033){\makebox(0,0)[lb]{\smash{{{\SetFigFont{10}{12.0}{\rmdefault}{\mddefault}{\updefault}$~q_Y \in B$~?}}}}}
\put(3000,633){\makebox(0,0)[lb]{\smash{{{\SetFigFont{10}{12.0}{\rmdefault}{\mddefault}{\updefault}$~q_N \in B$~?}}}}}
\path(1200,2133)(2850,3108)
\put(6000,3633){\makebox(0,0)[lb]{\smash{{{\SetFigFont{10}{12.0}{\rmdefault}{\mddefault}{\updefault}Reject}}}}}
\put(5250,108){\makebox(0,0)[rb]{\smash{{{\SetFigFont{7}{8.4}{\rmdefault}{\mddefault}{\updefault}no}}}}}
\put(6000,2433){\makebox(0,0)[lb]{\smash{{{\SetFigFont{10}{12.0}{\rmdefault}{\mddefault}{\updefault}Accept}}}}}
\put(6000,1233){\makebox(0,0)[lb]{\smash{{{\SetFigFont{10}{12.0}{\rmdefault}{\mddefault}{\updefault}Accept}}}}}
\put(6000,33){\makebox(0,0)[lb]{\smash{{{\SetFigFont{10}{12.0}{\rmdefault}{\mddefault}{\updefault}Reject}}}}}
\put(1950,2658){\makebox(0,0)[rb]{\smash{{{\SetFigFont{7}{8.4}{\rmdefault}{\mddefault}{\updefault}yes}}}}}
\put(1950,1083){\makebox(0,0)[rb]{\smash{{{\SetFigFont{7}{8.4}{\rmdefault}{\mddefault}{\updefault}no}}}}}
\put(5250,3633){\makebox(0,0)[rb]{\smash{{{\SetFigFont{7}{8.4}{\rmdefault}{\mddefault}{\updefault}yes}}}}}
\put(5175,2508){\makebox(0,0)[rb]{\smash{{{\SetFigFont{7}{8.4}{\rmdefault}{\mddefault}{\updefault}no}}}}}
\put(5250,1233){\makebox(0,0)[rb]{\smash{{{\SetFigFont{7}{8.4}{\rmdefault}{\mddefault}{\updefault}yes}}}}}
\end{picture}
}

     \end{center}
\caption{A possible behavior of $M(x)$
(applies to Section~\ref{s:PH-p} and Section~\ref{s:BH-p}).\label{fig:query1}} 

\end{figure}

On input $x$, $M'$  determines the first 
query of $M(x)$ and the two potential second
queries of $M(x)$.  We will write $q$ for the first 
query asked by $M(x)$, $q_{Y}$ for the second query that
would be asked by $M(x)$ 
were it to receive
a ``yes'' answer to the first query, and
$q_{N}$ for the second query that would be asked by $M(x)$
were it to receive a ``no'' answer to the first query.
$M'(x)$ then determines for which 
of the four possible answers to two sequential
queries (namely, ``no/no,'' 
``no/yes,'' 
``yes/no,'' and 
``yes/yes'')
$M(x)$ accepts.
All this can be done in polynomial time without querying any strings.
Since $M(x)$ asks two queries, there are sixteen different
possibilities for 
Accept/Reject behavior of $M(x)$.  As an example, suppose
that $M(x)$ accepts if and only if the
answer to the first query differs from the answer to the second query
(see Figure~\ref{fig:query1} for a pictorial representation of this case).

\begin{figure}[tp]
  \begin{center}

\setlength{\unitlength}{0.00050000in}
\begingroup\makeatletter\ifx\SetFigFont\undefined%
\gdef\SetFigFont#1#2#3#4#5{%
  \reset@font\fontsize{#1}{#2pt}%
  \fontfamily{#3}\fontseries{#4}\fontshape{#5}%
  \selectfont}%
\fi\endgroup%
{\renewcommand{\dashlinestretch}{30}
\begin{picture}(7684,3780)(0,-10)
\path(1200,1758)(3150,708)
\path(5700,933)(7350,1308)
\path(5700,558)(7350,108)
\path(5700,2958)(7350,2508)
\path(5700,3333)(7350,3708)
\put(3300,3033){\makebox(0,0)[lb]{\smash{{{\SetFigFont{10}{12.0}{\rmdefault}{\mddefault}{\updefault}$~q \in A \wedge q_Y \in B$~?}}}}}
\put(3300,633){\makebox(0,0)[lb]{\smash{{{\SetFigFont{10}{12.0}{\rmdefault}{\mddefault}{\updefault}$~q \in A \wedge q_Y \not\in B$~?}}}}}
\put(2250,2733){\makebox(0,0)[rb]{\smash{{{\SetFigFont{7}{8.4}{\rmdefault}{\mddefault}{\updefault}yes}}}}}
\path(1200,2133)(3150,3108)
\put(2250,1008){\makebox(0,0)[rb]{\smash{{{\SetFigFont{7}{8.4}{\rmdefault}{\mddefault}{\updefault}no}}}}}
\put(6525,183){\makebox(0,0)[rb]{\smash{{{\SetFigFont{7}{8.4}{\rmdefault}{\mddefault}{\updefault}no}}}}}
\put(0,1833){\makebox(0,0)[lb]{\smash{{{\SetFigFont{10}{12.0}{\rmdefault}{\mddefault}{\updefault}$q_N \in B$~?}}}}}
\put(7500,3633){\makebox(0,0)[lb]{\smash{{{\SetFigFont{10}{12.0}{\rmdefault}{\mddefault}{\updefault}Reject}}}}}
\put(7500,2433){\makebox(0,0)[lb]{\smash{{{\SetFigFont{10}{12.0}{\rmdefault}{\mddefault}{\updefault}Accept}}}}}
\put(7500,1233){\makebox(0,0)[lb]{\smash{{{\SetFigFont{10}{12.0}{\rmdefault}{\mddefault}{\updefault}Accept}}}}}
\put(7500,33){\makebox(0,0)[lb]{\smash{{{\SetFigFont{10}{12.0}{\rmdefault}{\mddefault}{\updefault}Reject}}}}}
\put(6525,3558){\makebox(0,0)[rb]{\smash{{{\SetFigFont{7}{8.4}{\rmdefault}{\mddefault}{\updefault}yes}}}}}
\put(6525,1158){\makebox(0,0)[rb]{\smash{{{\SetFigFont{7}{8.4}{\rmdefault}{\mddefault}{\updefault}yes}}}}}
\put(6525,2583){\makebox(0,0)[rb]{\smash{{{\SetFigFont{7}{8.4}{\rmdefault}{\mddefault}{\updefault}no}}}}}
\end{picture}
}

     \end{center}
\caption{Query structure of the $\p^{\np:\sigmatwo}$ machine corresponding
to Figure~\ref{fig:query1}.\label{fig:query2}}

\end{figure}
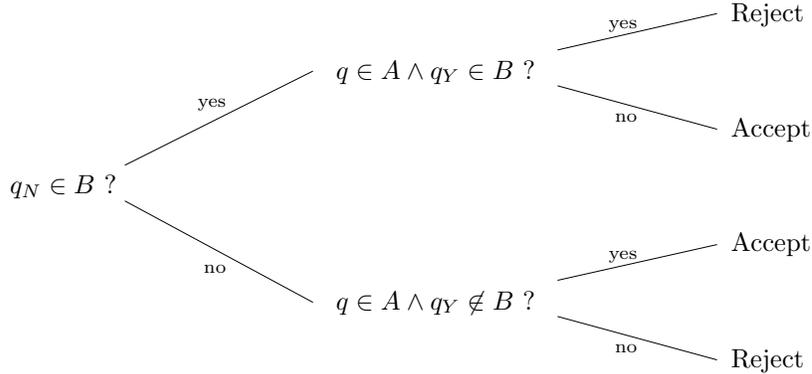

In this case, $M'(x)$ proceeds as shown in Figure~\ref{fig:query2}.
It is easy to see that $M'$ accepts $x$ if and only if $M$ accepts $x$.
In addition, note that the first query in 
Figure~\ref{fig:query2} is a query to 
an NP set, namely $B$,
and that the two potential second queries
``$q\in A \,\,\wedge\,\, q_Y \in B$?''~and
``$q\in A\,\,\wedge\,\, q_Y\not\in B$?''~are both
$\sigmatwo$ predicates. 
Since $\sigmatwo$ is closed under disjoint union we are done
for the case corresponding to Figure~\ref{fig:query1}.
There are of course fifteen other cases to consider, but
all of these are similar to or easier than the case we just treated.  We
mention in passing
that the full proof of Theorem~\ref{t:exchange}
in~\cite{hem-hem-hem:ctoappear:query-order-ph}
is more elegant than working through the sixteen different subcases
and splicing them together dynamically;  the 
present proof 
fragment is merely intended to convey some of the flavor of 
how one can prove things about query order.
\qed

\section{Query Order in the Boolean Hierarchy}\label{s:BH}

\subsection{Results}
\label{subs:BH-r-o}

The boolean hierarchy~\cite{cai-gun-har-hem-sew-wag-wec:j:bh1,cai-gun-har-hem-sew-wag-wec:j:bh2} was introduced in the 1980s, and captures and classifies 
those languages that can be computed via finite hardware operating over $\np$ 
predicates (equivalently, that can be computed via bounded access to SAT).

\begin{definition}
\begin{enumerate}
\item Let $\calc\ominus\cald=\{ L_1 - 
L_2 \condition  L_1 \in \calc \land 
L_2 \in \cald\}$.
\item\cite{cai-gun-har-hem-sew-wag-wec:j:bh1}\nopagebreak%
\begin{enumerate}%
\item $\bh_0=\p$.
\item For each $k>0$, $\bh_k=\np\ominus\bh_{k-1}$.
\item $\BH$, the boolean hierarchy, is defined as $\bigcup\limits_{k\geq 0}\bh_k$. 
\end{enumerate}
\end{enumerate}
\end{definition}

So $\bh_1=\np$, and $\bh_2$ equals Papadimitriou and 
Yannakakis's~\cite{pap-yan:j:dp} class DP, namely the class of all sets that 
can be written as the intersection of some $\np$ set with some $\conp$ set.
NP, DP, and the other levels of the boolean hierarchy contain a large 
variety of complete problems 
(see, 
e.g.,~\cite{gar-joh:b:int,cai-mey:j:dp,cai-gun-har-hem-sew-wag-wec:j:bh1,bor:t:complexity-of-mind-changes}). 

Hemaspaandra, Hempel, and 
Wechsung~\cite{hem-hem-wec:tSPECIALwithJ:query-order-bh} raised the topic of query order by asking whether $\pij=\pji$.
They resolved this question as follows.
They noted that equality trivially holds if $i=j  \,\,\lor\,\,  i=0  \,\,\lor\,\,  j=0$.
They proved that equality (not so trivially) holds if 
$i \mbox{ is even}\,\,\land \,\, j=i+1$,
or if $j \mbox{ is even} \,\,\land\,\, i=j+1$.
They proved that for all other cases inequality holds unless the polynomial 
hierarchy collapses. 

\begin{figure}[tp]
\begin{center}
$$i\quad\left.
\begin{minipage}{10cm}
\begin{tabular}{|c||c|c|c|c|c|c|c|c|c}
:&etc.&etc.&etc.&etc.&etc.&etc.&etc.&etc.&etc.\\ \hline
7&\mbox{\Large\mathversion{bold}$=$}&\substar&\substar&\substar&\substar&\substar&\mbox{\Large\mathversion{bold}$=$}&\mbox{\Large\mathversion{bold}$=$}&etc.\\ \hline
6&\mbox{\Large\mathversion{bold}$=$}&\substar&\substar&\substar&\substar&\substar&\mbox{\Large\mathversion{bold}$=$}&\mbox{\Large\mathversion{bold}$=$}&etc.\\ \hline
5&\mbox{\Large\mathversion{bold}$=$}&\substar&\substar&\substar&\mbox{\Large\mathversion{bold}$=$}&\mbox{\Large\mathversion{bold}$=$}&\superstar&\superstar&etc.\\ \hline
4&\mbox{\Large\mathversion{bold}$=$}&\substar&\substar&\substar&\mbox{\Large\mathversion{bold}$=$}&\mbox{\Large\mathversion{bold}$=$}&\superstar&\superstar&etc.\\ \hline
3&\mbox{\Large\mathversion{bold}$=$}&\substar&\mbox{\Large\mathversion{bold}$=$}&\mbox{\Large\mathversion{bold}$=$}&\superstar&\superstar&\superstar&\superstar&etc.\\ \hline
2&\mbox{\Large\mathversion{bold}$=$}&\substar&\mbox{\Large\mathversion{bold}$=$}&\mbox{\Large\mathversion{bold}$=$}&\superstar&\superstar&\superstar&\superstar&etc.\\ \hline
1&\mbox{\Large\mathversion{bold}$=$}&\mbox{\Large\mathversion{bold}$=$}&\superstar&\superstar&\superstar&\superstar&\superstar&\superstar&etc.\\ \hline
0&\mbox{\Large\mathversion{bold}$=$}&\mbox{\Large\mathversion{bold}$=$}&\mbox{\Large\mathversion{bold}$=$}&\mbox{\Large\mathversion{bold}$=$}&\mbox{\Large\mathversion{bold}$=$}&\mbox{\Large\mathversion{bold}$=$}&\mbox{\Large\mathversion{bold}$=$}&\mbox{\Large\mathversion{bold}$=$}&etc.\\ \hline\hline
&0&1&2&3&4&5&6&7&$\cdot\cdot$\\ \hline
\end{tabular}
$$j$$
\end{minipage}
\right.
$$

\label{table:1}
\caption{\label{fig:tablefig}Relationship between $\pij$ and $\pji$.}
\vspace{.2cm}
\begin{footnotesize}
\begin{tabular}{lcp{7cm}}
{\em Key:\/} &\mbox{\Large\mathversion{bold}$=$}~ in row $i$ and column $j$& means $\pij=\pji$.\\
&\substar~ in row $i$ and column $j$& means $\pij$ is a strict subset of 
$\pji$ unless the polynomial hierarchy collapses.\\
&\superstar~ in row $i$ and column $j$& means $\pij$ is a strict superset of 
$\pji$ unless the polynomial hierarchy collapses.
\end{tabular}
\end{footnotesize}
\end{center}
\end{figure}

\begin{theorem}\label{t:HHW-xf}~
\cite{hem-hem-wec:tSPECIALwithJ:query-order-bh} \quad
For each $i$ and $j$, the 
relationship between $\pij$ and $\pji$ is as shown in 
Figure~\ref{fig:tablefig}. 
\end{theorem}

The most strikingly odd feature of this theorem is that the just-off-diagonal 
entries alternate between equality and inequality (e.g., 
$\p^{\bh_2:\bh_3}=\p^{\bh_3:\bh_2}$, yet 
unless the polynomial hierarchy collapses
$\p^{\bh_3:\bh_4}\not=\p^{\bh_4:\bh_3}$).
The curious asymmetry becomes a bit less opaque if one looks at what is 
actually underpinning Theorem~\ref{t:HHW-xf}.
The key result on which Theorem~\ref{t:HHW-xf} rests is Lemma~\ref{l:zz} 
below, which states that ordered access to the boolean hierarchy's levels 
can without loss of generality be restructured as parallel access to $\np$. 
As is standard, let ${\rm R}^{\littlep}_{\ell\mbox{{\scriptsize -tt}}}(\np)$ denote 
$\{L\condition(\exists A \in \np)[L\leq^{\littlep}_{\ell\mbox{{\scriptsize -tt}}}A]\}$,
  where 
$\leq^{\littlep}_{\ell\mbox{{\scriptsize -tt}}}$ is the standard reduction allowing 
$\ell$ parallel queries~\cite{lad-lyn-sel:j:com}.

\begin{lemma}\label{l:zz}
For each $i,j\geq 1$,
$$
\pij=
\left\{
\begin{array}{l@{\qquad}p{6cm}}
{\rm R}^{\littlep}_{(i+2j-1)\mbox{\scriptsize -tt}}(\np) & if $i$ is even and $j$ is odd\\
{\rm R}^{\littlep}_{(i+2j)\mbox{\scriptsize -tt}}(\np) & otherwise.
\end{array}
\right.
$$
\end{lemma}

This lemma is the source of the asymmetry between,
for example, 2-versus-3 and 3-versus-4. 
Of course, this in some way begs 
the question, as the reader may well ask about the asymmetry between 
the first and second cases of Lemma~\ref{l:zz}. 
Briefly and informally put, when $i$ is even and $j$ is odd, a certain 
underlying graph modeling the computation of $\pij$ becomes non-bipartite, 
and by doing so allows one to guarantee a savings of one parallel query to 
$\np$ 
(see~\cite{hem-hem-wec:tSPECIALwithJ:query-order-bh} for full details).

\subsection{Proof by Example}\label{s:BH-p}

In this subsection
we will give a partial proof for an instance of 
Lemma~\ref{l:zz}, namely, we  
will give a partial proof for
$$\p^{{\rm DP}:\np} 
\subseteq {\rm R}^{\littlep}_{3\mbox{\scriptsize -tt}}(\np).$$
In particular, we will show that 
$X
\subseteq {\rm R}^{\littlep}_{3\mbox{\scriptsize -tt}}(\np)$,
where $X$ is the 
class of 
languages that are in 
$\p^{{\rm DP}:\np}$
via a 
$\p^{{\rm DP}:\np}$
machine in which the~$\p$ machine 
on each input asks {\em exactly\/} one question to each of 
its oracles, and in which the~$\p$ machine accepts 
if and only if {\em exactly one of its two queries gets
the answer ``yes.''}  That is, as was the 
case also in Section~\ref{s:PH-p}, we will do the parity case.

We warn the reader that
the proof approach taken here is 
not suited to be elegantly
generalized to eventually yield Lemma~\ref{l:zz}. 
For a complete 
and unified proof of $\p^{{\rm DP}:\np} \subseteq 
{\rm R}^{\littlep}_{3\mbox{\scriptsize -tt}}(\np)$, we refer 
the reader to~\cite{hem-hem-wec:tSPECIALwithJ:query-order-bh}. 
However, the proof given below provides a good starting point for 
understanding how these proofs work in the context of the boolean hierarchy.
Note that the proof picks one of the interesting 
cases of 
Lemma~\ref{l:zz}
(``$i$~even and $j$~odd'').  As hands-on experience is the best 
way to get a feel for an area, we urge the reader to 
come up with his or her own proof before reading the proof below.

\noindent{\bf Proof:}
Let $L \in \p^{{\rm DP}:\np}$. Let~$\p$ machine~$M$,  
$A\in {\rm DP}$,  and $B \in \np$ be such that $M$ accepts $L$ and, 
on each input, $M$
makes exactly one query to $A$ followed by 
exactly one query to $B$.
We will partially describe a $\p$ machine $M'$ that accepts
$L$ with three parallel queries to an $\np$ set. 

On input $x$, $M'$  determines the first 
query and the two potential second
queries of $M(x)$.
As in the proof of Section~\ref{s:PH-p}, we will
write $q$ for the first
query asked by $M(x)$, $q_{Y}$ for the second query 
that would be asked by $M(x)$
were it to
receive a ``yes'' answer to the first query, and
$q_{N}$ for the second query that would be asked by $M(x)$
were it to receive a ``no'' answer to the first query.
$M'(x)$ 
then determines for which answers to its two queries $M(x)$ accepts.
All this can be done in polynomial time without querying any strings.
Again, we will consider the case pictured in
Figure~\ref{fig:query1}, i.e., the case 
that $M$ accepts $x$ if and only if the
answer to the first query differs from the answer to the second query.

Let $A_1, A_2 \in \np$ be such that $A=A_1-A_2$. 
It is well-known that we may choose $A_1$ and 
$A_2$ to be such that
$A_1\superseteq A_2$~\cite{cai-gun-har-hem-sew-wag-wec:j:bh1}.
If we know the answers to the {\em four\/} $\np$
queries, 
$q \in A_1$, 
$q \in A_2$,
$q_N \in B$, and
$q_Y \in B$,
we can easily determine whether $M(x)$ accepts or rejects.
Thus we have  $\p^{{\rm DP}:\np} 
\subseteq {\rm R}^{\littlep}_{4\mbox{\scriptsize -tt}}(\np)$. 
How can we do better? In particular, how can we save one parallel $\np$ query? 

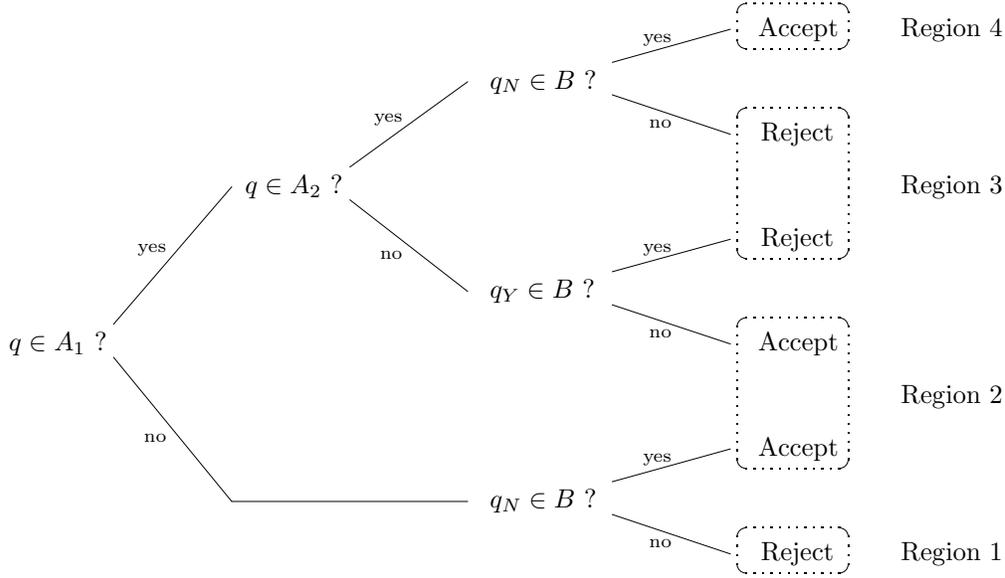
\begin{figure}[tp]
\setlength{\unitlength}{0.00046000in}
\begingroup\makeatletter\ifx\SetFigFont\undefined%
\gdef\SetFigFont#1#2#3#4#5{%
  \reset@font\fontsize{#1}{#2pt}%
  \fontfamily{#3}\fontseries{#4}\fontshape{#5}%
  \selectfont}%
\fi\endgroup%
{\renewcommand{\dashlinestretch}{30}
\begin{picture}(9612,6564)(0,-10)
\path(6900,1062)(8250,1437)
\path(6900,3087)(8250,2637)
\path(6900,3462)(8250,3837)
\path(6900,5487)(8250,5037)
\path(6900,5862)(8250,6237)
\path(3900,4662)(5250,5637)
\path(3900,4287)(5250,3237)
\path(2550,837)(5250,837)
\path(1200,2862)(2550,4437)
\path(1200,2487)(2550,837)
\put(8430,117){\arc{210}{1.5708}{3.1416}}
\put(8430,432){\arc{210}{3.1416}{4.7124}}
\put(9495,432){\arc{210}{4.7124}{6.2832}}
\put(9495,117){\arc{210}{0}{1.5708}}
\dottedline{101}(8325,117)(8325,432)
\dottedline{101}(8430,537)(9495,537)
\dottedline{101}(9600,432)(9600,117)
\dottedline{101}(9495,12)(8430,12)
\put(8430,1317){\arc{210}{1.5708}{3.1416}}
\put(8430,2832){\arc{210}{3.1416}{4.7124}}
\put(9495,2832){\arc{210}{4.7124}{6.2832}}
\put(9495,1317){\arc{210}{0}{1.5708}}
\dottedline{101}(8325,1317)(8325,2832)
\dottedline{101}(8430,2937)(9495,2937)
\dottedline{101}(9600,2832)(9600,1317)
\dottedline{101}(9495,1212)(8430,1212)
\put(8430,3717){\arc{210}{1.5708}{3.1416}}
\put(8430,5232){\arc{210}{3.1416}{4.7124}}
\put(9495,5232){\arc{210}{4.7124}{6.2832}}
\put(9495,3717){\arc{210}{0}{1.5708}}
\dottedline{101}(8325,3717)(8325,5232)
\dottedline{101}(8430,5337)(9495,5337)
\dottedline{101}(9600,5232)(9600,3717)
\dottedline{101}(9495,3612)(8430,3612)
\put(8430,6117){\arc{210}{1.5708}{3.1416}}
\put(8430,6432){\arc{210}{3.1416}{4.7124}}
\put(9495,6432){\arc{210}{4.7124}{6.2832}}
\put(9495,6117){\arc{210}{0}{1.5708}}
\dottedline{101}(8325,6117)(8325,6432)
\dottedline{101}(8430,6537)(9495,6537)
\dottedline{101}(9600,6432)(9600,6117)
\dottedline{101}(9495,6012)(8430,6012)
\put(0,2562){\makebox(0,0)[lb]{\smash{{{\SetFigFont{10}{12.0}{\rmdefault}{\mddefault}{\updefault}$q \in A_1$~?}}}}}
\put(2700,4362){\makebox(0,0)[lb]{\smash{{{\SetFigFont{10}{12.0}{\rmdefault}{\mddefault}{\updefault}$q \in A_2$~?}}}}}
\put(5400,762){\makebox(0,0)[lb]{\smash{{{\SetFigFont{10}{12.0}{\rmdefault}{\mddefault}{\updefault}$~q_{N} \in B$~?}}}}}
\path(6900,687)(8250,237)
\put(8370,1362){\makebox(0,0)[lb]{\smash{{{\SetFigFont{10}{12.0}{\rmdefault}{\mddefault}{\updefault}~~Accept}}}}}
\put(7575,312){\makebox(0,0)[rb]{\smash{{{\SetFigFont{7}{8.4}{\rmdefault}{\mddefault}{\updefault}no}}}}}
\put(8400,162){\makebox(0,0)[lb]{\smash{{{\SetFigFont{10}{12.0}{\rmdefault}{\mddefault}{\updefault}~~Reject}}}}}
\put(8370,2562){\makebox(0,0)[lb]{\smash{{{\SetFigFont{10}{12.0}{\rmdefault}{\mddefault}{\updefault}~~Accept}}}}}
\put(8400,3762){\makebox(0,0)[lb]{\smash{{{\SetFigFont{10}{12.0}{\rmdefault}{\mddefault}{\updefault}~~Reject}}}}}
\put(5400,3162){\makebox(0,0)[lb]{\smash{{{\SetFigFont{10}{12.0}{\rmdefault}{\mddefault}{\updefault}$~q_{Y} \in B$~?}}}}}
\put(8370,6162){\makebox(0,0)[lb]{\smash{{{\SetFigFont{10}{12.0}{\rmdefault}{\mddefault}{\updefault}~~Accept}}}}}
\put(8400,4962){\makebox(0,0)[lb]{\smash{{{\SetFigFont{10}{12.0}{\rmdefault}{\mddefault}{\updefault}~~Reject}}}}}
\put(5400,5562){\makebox(0,0)[lb]{\smash{{{\SetFigFont{10}{12.0}{\rmdefault}{\mddefault}{\updefault}$~q_{N} \in B$~?}}}}}
\put(4500,5187){\makebox(0,0)[rb]{\smash{{{\SetFigFont{7}{8.4}{\rmdefault}{\mddefault}{\updefault}yes}}}}}
\put(4500,3612){\makebox(0,0)[rb]{\smash{{{\SetFigFont{7}{8.4}{\rmdefault}{\mddefault}{\updefault}no}}}}}
\put(1800,3687){\makebox(0,0)[rb]{\smash{{{\SetFigFont{7}{8.4}{\rmdefault}{\mddefault}{\updefault}yes}}}}}
\put(1800,1512){\makebox(0,0)[rb]{\smash{{{\SetFigFont{7}{8.4}{\rmdefault}{\mddefault}{\updefault}no}}}}}
\put(7575,6087){\makebox(0,0)[rb]{\smash{{{\SetFigFont{7}{8.4}{\rmdefault}{\mddefault}{\updefault}yes}}}}}
\put(7575,5112){\makebox(0,0)[rb]{\smash{{{\SetFigFont{7}{8.4}{\rmdefault}{\mddefault}{\updefault}no}}}}}
\put(7575,3687){\makebox(0,0)[rb]{\smash{{{\SetFigFont{7}{8.4}{\rmdefault}{\mddefault}{\updefault}yes}}}}}
\put(7575,2712){\makebox(0,0)[rb]{\smash{{{\SetFigFont{7}{8.4}{\rmdefault}{\mddefault}{\updefault}no}}}}}
\put(7575,1287){\makebox(0,0)[rb]{\smash{{{\SetFigFont{7}{8.4}{\rmdefault}{\mddefault}{\updefault}yes}}}}}

\put(9900,6162){\makebox(0,0)[lb]{\smash{{{\SetFigFont{10}{12.0}{\rmdefault}{\mddefault}{\updefault}~~~Region 4}}}}}
\put(9900,4362){\makebox(0,0)[lb]{\smash{{{\SetFigFont{10}{12.0}{\rmdefault}{\mddefault}{\updefault}~~~Region 3}}}}}
\put(9900,1962){\makebox(0,0)[lb]{\smash{{{\SetFigFont{10}{12.0}{\rmdefault}{\mddefault}{\updefault}~~~Region 2}}}}}
\put(9900,162){\makebox(0,0)[lb]{\smash{{{\SetFigFont{10}{12.0}{\rmdefault}{\mddefault}{\updefault}~~~Region 1}}}}}
\end{picture}
}

   \caption{Refined query tree of $M(x)$---all queries drawn are $\np$
queries.} 
  \label{fig:query3}
\end{figure}

Let us redraw the query tree of $M(x)$ using the 
above-mentioned underlying 
four queries in the fashion 
shown in Figure~\ref{fig:query3}.   
Recall that 
we are looking at the case in which $M$ accepts $x$ if and only if
the
answer to the first query differs from the answer to the second query.
Since $A_1 \superseteq A_2$, if $q\not\in A_1$ then 
$q \not \in A$.

The refined query tree of $M(x)$ displays 
four regions of acceptance and rejection 
(see Figure~\ref{fig:query3}). 
In order to correctly simulate $M(x)$,  it suffices 
to find out in which region the correct branch 
ends. 
However, this can be done with just three questions, namely:
\begin{enumerate}
\item Does the correct branch end in region 2, 3, or 4?
($q \in A_1 \,\,\vee\,\, q_N\in B$?) 
\item Does it end in region 3 or 4? 
($q \in A_1 \,\,\wedge\,\, ( q \in A_2 \,\,\vee\,\, q_Y\in B)$?)
\item Does the correct branch end in region 4?
($q \in A_1 \,\,\wedge\,\, q \in A_2 \,\,\wedge\,\, q_N\in B$?)
\end{enumerate}

The answers to these three questions determine 
the region in which the 
correct branch ends and hence we know whether 
$M(x)$ rejects or accepts. 
In particular, $M'(x)$ 
should accept if and only if the correct branch ends
either in region 2 or region 4 (that is,
if and only if either only question (1)
is answered ``yes'' or all three questions are answered ``yes''). Note that we
use three different $\np$ sets and also various ``and''s and ``or''s in the
above description of the questions, but 
since $\np$ is closed under union, intersection, and disjoint union,
the three questions can be transformed (in polynomial time) into
three single queries that in turn can be asked (in parallel) to one $\np$ set.
\qed

\section{Related Work}

Sections~\ref{s:PH} and~\ref{s:BH} presented the 
basic results known about query 
order in the polynomial and boolean hierarchies. 
In a nutshell, query order never matters in the polynomial hierarchy, and in 
the boolean hierarchy we know in exactly which cases query order matters 
(assuming that the polynomial hierarchy does not collapse). 

However, via the study of query order, a number of results have been obtained 
regarding topics that at first blush might seem totally unrelated, such as 
bottleneck computation and downward translation of equality. 
Also, a number of researchers have generalized from ``one query to a given 
class'' to more elaborate settings such as tree-like query structures, 
multiple queries, and multiple rounds of multiple queries. 
In this section, we briefly provide pointers to these related topics and 
generalized settings.

\subsubsection*{Translating Equalities Downwards}

Suppose two questions to $\sigmak$ yield no new languages beyond those 
already solvable via one query to $\sigmak$. 
What follows?
 
Until very recently,
all one could conclude from this 
assumption was 
that the polynomial hierarchy collapses to a level slightly above  
one query to $\Sigma^{\littlep}_{k+1}$ (note the 
``+1'' 
here)~\cite{kad:joutdatedbychangkadin:bh,wag:t:n-o-q-87version,wag:t:n-o-q-89version,cha-kad:j:closer,bei-cha-ogi:j:difference-hierarchies}.
However,
work growing directly out of the study of query-order classes---namely, out of 
the goal of showing that different ordered access to levels of the polynomial 
hierarchy yields different language classes (see Theorem~\ref{t:mpxx})---led 
to a collapse a full level lower in the polynomial hierarchy. 
In particular, the one-two punch of~\cite{hem-hem-hem:tOutByConf:downward-translation,hem-hem-hem:jtoappear:downward-translation}/\cite{buh-for:t:two-queries}
yielded the following theorem. 

\begin{theorem}\label{t:one}
Let $k>1$. If $\p^{\sigmak[1]}=\p^{\sigmak[2]}$, then 
$\ph=\sigmak$.
\end{theorem}
In other words, if $\p^{\sigmak[1]}=\p^{\sigmak[2]}$, $k>1$, the polynomial 
hierarchy crashes to a class that (before the crash) was 
seemingly even lower than that 
at which the hypothesis's equality stands.

It has been 
shown~\cite{hem-hem-hem:jtoappear:downward-translation,hem-hem-hem:t:translating-downwards} 
that the above theorem in fact has an analog 
for the $j$ versus $j+1$ queries case. In particular,
we have the following, which like Theorem~\ref{t:one} 
was established in the literature 
via proving a more general theorem about 
query-order classes, and then deriving the stated result as 
a corollary to the more general 
theorem~\cite{hem-hem-hem:t:translating-downwards}.

\begin{theorem}\label{t:j}
Let $k>1$ and $m\geq 1$. If $\p^{\sigmak[m]}=\p^{\sigmak[m+1]}$, then 
${\rm DIFF}_m(\sigmak)={\rm co DIFF}_m(\sigmak)$.
\end{theorem}
Again, this says that, under the stated assumption, there 
is a collapse within the boolean hierarchy to a level that, 
{\em a priori}, was just {\em below\/} $\p^{\sigmak[m]}$.

In a nutshell, in this setting smaller classes collapse if and only if 
larger classes collapse---a type of behavior people have been stalking ever 
since influential papers of Book~\cite{boo:j:tally} and Hartmanis, 
Immerman, and 
Sewelson~\cite{har:j:sparse,har-imm-sew:j:sparse}
raised the issue of when classes stand and fall together.

\subsubsection*{Multiple Queries and Bottleneck Computations}

In this survey, we have focused on the most natural case: one query to each 
of the two information sources. 
A number of papers building on those mentioned here have studied more 
elaborate settings. 

In fact, 
the initial paper of Hemaspaandra, Hempel, and 
Wechsung~\cite{hem-hem-wec:tSPECIALwithJ:query-order-bh} already 
studied the case of general tree-like access 
to levels of the boolean hierarchy, and in doing so 
studied the case of multiple rounds of single queries; 
Beigel and Chang~\cite{bei-cha:c:commutative-queries} 
study multiple rounds of multiple 
queries to the polynomial hierarchy, 
and show that here the order does not matter, and they also study 
the case of function classes; 
Wagner~\cite{wag:t:parallel-difference} studies 
parallel rounds of one or more queries to the polynomial hierarchy and other 
classes and 
also tightly relates such classes 
to the refined hierarchy work of 
Selivanov~\cite{sel:c:refined-ph,sel:j:fine-hierarchies}
(see also the discussion in 
the final paragraph of Section 2 
of~\cite{hem-hem-hem:cSpecialHardwiredConferenceCite:sn-1tt-np-completeness}).

In a quite different direction, bottleneck machines are a model used
to study whether a computational problem can be decomposed into a
large number of simple, sequential, tasks, each of which passes on
only a very limited amount of information to the next task, and all of
which differ only in that input and in a ``task number''
input~\cite{cai-fur:j:bottleneck}.  A surprising recent paper of
Hertrampf~\cite{her:c:transformation-monoid-acceptance} uses ordered
access involving multiple queries, combined with quantifier-based and
modulo-based computation, to completely characterize the languages
accepted by certain bottleneck machine classes---classes that had long
eluded crisp characterization.

\subsubsection*{Advice Classes, Self-Specifying Machines, and Completeness Types}

A number of other seemingly different notions are related to query order. 
Hemaspaandra, Hempel, and Wechsung~\cite{hem-hem-wec:t2:self-specifying} have 
studied self-specifying machines---nondeterministic machines that dynamically 
specify the path sets on which they will accept. 
They completely characterize the two most natural such classes in terms of 
query-order classes with a ``positive final query'' restriction.
They show that the classes have equivalent characterizations as the 
$\sharpp$-closures of P and NP, respectively, and they establish a query 
order result mixing function and language classes:
$
\p^{\sharpp[1]}=\p^{\sharpp:\np}\iff 
\p^{\sharpp[1]}=\p^{\sharpp:\np[{\cal O}(1)]}
$
(where ``$\calc:\cald$'' access means one query each except when 
${\cal O}(1)$ queries are explicitly stated for the queried class). 
They also show that the classes have
characterizations in terms 
of the ``input-specific advice'' notation of K\"obler and 
Thierauf~\cite{koe-thi:j:opt}.

Agrawal, Beigel, and Thierauf~\cite{agr-bei-thi:tc:modulo-information}, 
independently 
of~\cite{hem-hem-wec:tSPECIALwithJ:query-order-bh}, also study 
input-specific-advice classes. 
As noted 
by 
Hemaspaandra et al.~in 
the journal version of~\cite{hem-hem-wec:tSPECIALwithJ:query-order-bh}, 
this can be seen as equivalent to studying query order with a ``positive 
final query'' restriction---i.e., the machines must ``do'' exactly what the 
response to their second query is. 
A detailed 
and careful discussion of the relationship between the two papers 
can be found 
in the journal version of~\cite{hem-hem-wec:tSPECIALwithJ:query-order-bh}.

Finally, a long line of research has asked whether $\manyone$-completeness 
and $\Turing$-completeness stand or fall
together for classes that 
potentially lack complete sets.
Gurevich~\cite{gur:c:comp} and Ambos-Spies~\cite{amb:j:complete-problems} 
have shown that, for all classes $\calc$ closed downwards under Turing 
reductions, it holds robustly that: $\calc$ has $\manyone$-complete sets if 
and only if $\calc$ has $\Turing$-complete sets. 
Nonetheless, by studying a strong nondeterministic closure of NP
that, it turns out, exactly equals the query-order class 
$\p^{\np\cap\conp:\np}$, 
Hemaspaandra et al.~have recently shown that on some reducibility 
closures of NP, $\manyone$-completeness and $\Turing$-completeness do not 
robustly stand or fall 
together~\cite{hem-hem-hem:cSpecialHardwiredConferenceCite:sn-1tt-np-completeness}.

\singlespacing
\bibliography{gry}

\end{document}